\documentclass{ws-ijmpa}
\usepackage[super,compress]{cite}
\usepackage{graphicx}
\usepackage{xcolor}
\usepackage{slashed}
\begin{document}
\markboth{Garc\'ia-Aguilar J.D., P\' erez-Lorenzana A.}{Implications of Lorentz symmetry violation on a 5D supersymmetric 
model}

\catchline{}{}{}{}{}
%
%%%%%%%%%%%%%%%%%%%%%%%%%%%%%%%%%%%%%%%%%%%%%%%%%%%%%%%%%%%%%%%%%%%%

\title{Implications of Lorentz symmetry violation on a 5D supersymmetric 
model
}

\author{Garc\'ia-Aguilar J. D.
}

%\address{University Department, University Name, Address\\
%City, State ZIP/Zone, Country\footnote{
%State completely without abbreviations, the affiliation and
%mailing address, including country. Typeset in 8 pt italic.}\\
\address{jdgarcia@fis.cinvestav.mx}

\author{P\'erez-Lorenzana A.}

\address{aplorenz@fis.cinvestav.mx\\
Departamento de F\'isica, Centro de Investigaci\'on y de Estudios Avanzados del I.P.N.. 
Apdo. Post. 14740, 0700, Mexico City, M\'exico}

\maketitle

\begin{history}
\received{Day Month Year}
\revised{Day Month Year}
\end{history}

\begin{abstract}
Field models with $n$ extra spatial dimensions have a larger $SO(1,3+n)$ 
Lorentz symmetry which is broken down to the standard $SO(1,3)$ four 
dimensional one by the compactification process. By considering  
Lorentz violating operators in a $5D$ supersymmetric Wess-Zumino mo\-del, which 
otherwise conserve the standard four dimensional Poincare invariance, we show 
that supersymmetry can be restored upon a simple deformation of the 
supersymmetric transformations. However, 
supersymmetry is not preserved in the effective $4D$ theory that arises 
after compactification when the $5D$ Lorentz violating operators do not 
preserve  $Z_2: y\rightarrow -y$ bulk parity. 
Our mechanism unveils a possible connection among Lorentz violation and 
the Scherk-Schwarz mechanism.
We also show that parity preserving models, on the other hand, do provide well 
defined supersymmetric KK models.

\end{abstract}

\ccode{PACS numbers: 11.10.Kk, 12.60.Jv, 04.50.Cd, 11.30.Cp, 11.30.Pb}

%\tableofcontents

\section{Introduction}	
Symmetries play an important role to describe the fundamental interactions on particle physics 
\cite{coleman1988aspects} in the context of the Standard Model (SM). Those  can be 
categorized in two general sets, the so called internal and 
the space-time symmetries. Internal symmetries do not affect space-time coordinates, 
but are responsible of implementing the gauge principle in quantum field 
theories. 
Space-time symmetries, on the other hand, are useful to label fundamental field 
properties. In the SM, the internal  $SU(3)_C\times SU(2)_L\times 
U(1)_Y$  and Lorentz symmetries can not be mixed into a single Lie group, because it is not possible to 
find a Lie Algebra which contains, in a non trivial way, the generators of both  
groups. 
This is possible, however, in supersymmetric theories \cite{Martin:1997ns}, 
which opens the possibility of finding a theory 
that incorporates the SM interactions and gravity on 
the same theoretical ground. One of the best candidates for that is String 
Theory, which needs supersymmetry to be self consistent. As a plus, 
supersymmetry 
alleviates some of the theoretical problems of the SM, as the hierarchy 
problem. On the other hand, String Theory also predicts additional 
space-like dimensions to the four we see. This last feature had motivated a lot 
of interest in the past decade for the study of models with extra 
dimensions, also because they may provide another approach to solve the 
hierarchy problem~\cite{XD,PerezLorenzana:2005iv}. Nevertheless, none of these 
had found so far an experimental confirmation, which indicates that the theory 
should have a way of hiding the physical implications of both, supersymmetry and 
extra dimensions. Last is usually explained from the condition that such new 
dimensions should be rather compact and of very small sizes, whereas  supersymmetry 
should be a broken symmetry at low energy. The actual physical mechanisms 
beneath those conditions are yet unknown. Exploring the theoretical potential of both ideas 
still remains as an interesting topic. We will be dealing with it along the present 
work.

Supersymmetry (SUSY) involves the existence of transformations between 
bosonic and 
fermionic fields whose Lie graded algebra in four dimensions is given by
\begin{equation}
\left[P_\mu, P_\nu\right]=\left[P_\mu, Q\right]=0,
\hspace{1cm}\lbrace Q,\bar{Q}\rbrace\propto 2\gamma^\mu P_\mu, \label{susytrans4D}
\end{equation}
where $P_\mu$ represents the energy-momentum vector which generates 
translations in 
space-time. Also, the $Q$ spinor generates the supersymmetric transformations and 
$\gamma^\mu$ are the Dirac matrices. 
For space-time dimensions other than four, the Lie 
graded algebra is preserved, but properties of 
the spinors and  gamma matrices may change~\cite{VanProeyen:1999ni}.

It is important to remark that the supersymmetric algebra (\ref{susytrans4D}) does not involve 
the generators of the Lorentz group (SO(1,n)), which allows to violate Lorentz invariance, at least in an explicit 
way, while preserving the  mathematical form 
of the Lie graded algebra, that is, preserving SUSY. In Ref. 
\citen{Berger:2001rm}, Kostelecky and Berger 
presented a 4D Wess-Zumino supersymmetric model which also contains specific operators that 
explicitly break Lorentz invariance~\cite{Colladay:1996iz}. There, they have 
shown that it is possible to  have SUSY in a Lorentz violating theory, provided 
SUSY transformations between the fields of the model are 
modified by the simple replacement rule given by
\begin{equation}
p_\mu\gamma^\mu\rightarrow p_\mu\left(\gamma^\mu+k_{\mu\nu}\gamma^\nu\right),
\end{equation}
where $k_{\mu\nu}$ is a set of parameters which could be seen as the non zero expectation value of a background 
tensorial field, which, therefore, quantify the violation of the Lorentz symmetry~\cite{Kostelecky:1988zi}.
It is worth noticing that the usual SUSY transformations for the Wess-Zumino model~\cite{Wess:1973kz} 
are recovered in the limit $k_{\mu\nu}\rightarrow 0$. 
A number of experiments have been used to impose strong constraints on 
many possible Lorentz violation parameters. 
A review with different strong limits can be found in Ref.~\citen{Kostelecky:2008ts}. 
Notice also that even though the model in 
Ref.~\citen{Berger:2001rm} is not considered a free (toy) model, 
the addition of Lorentz violation operators does not break SUSY, and  
therefore the model remains quite unviable to be a good phenomenological 
model, unless another mechanism for SUSY breaking is claimed~\cite{petrov}.

On the other hand, in the study of field theory models in more 
than four dimensions,  in order to understand why we see only three space-like 
dimensions, it is usually assumed that all extra dimensions are compactified on 
a closed space manifold of finite size \cite{PerezLorenzana:2005iv}.
However, the compactification process necessarily 
breaks the extended Lorentz symmetry, $SO(1,3+n)$, in an implicit way.
That is because any compactification has, 
by definition, the role of introducing a way to distinguish our four space-time 
dimensions from those extra ones. 
Therefore, no transformation that mixes normal with extra dimensions 
may remain as a symmetry after compactification. 
That is clearly the case, for instance, when compactification on  
orbifolds is used, where fixed po\-ints are introduced, or, either, when fields 
are located on the so called 4D branes.  
This implicit rupture of the higher dimensional symmetry is given only 
over the additional dimensions, 
because the resulting 4D effective model has the 
usual $4D$ Lorentz invariance. 
In formal terms, the compactification breaks the $SO(1,3+n)$ 
group into a residual $SO(1,3)$ group that only involves transformations among 
our standard four dimensions, such that Poincare invariance is always preserved 
at the level of the effective four dimensional theory.
On this line of reasoning, it seems interesting to consider the theoretical 
possibilities of  
those operators that explicitly break the larger Lorentz invariance, but which 
at the same time  preserve standard $4D$ Poincare symmetry.
The idea has been already explored by  Rizzo in Ref.~\citen{Rizzo:2005um}, where he has shown how some specific 
operators in $5D$ space-time models can achieve an explicit rupture of $5D$ 
Lorentz symmetry, but after  
compactification the effective model remains invariant under the corresponding 
$4D$ symmetry. The phenomenological consequ\-ences of these toy models were 
also discussed in there. However, a supersymmetric equivalent of this setup had 
not been 
considered so far.

In this work we address two related questions. First of all, we will explore 
whether or not SUSY can be restored in a $5D$ field theory where bulk Lorentz 
invariance is explicitly broken, but where the involved operators do conserve $4D$ Lorentz symmetry.  
Second, we will focus on the 
phenomenological consequences of this approach at the level of the effective (compactified) models.
Notice that in some cases it is possible to achieve supersymmetry breaking 
through the addition of operators over the free model \cite{Marti:2001iw}. 
For these purposes we will follow the 4D proposal on Ref.~\citen{Berger:2001rm},
but here, we will consider a $5D$  supersymmetric 
model where some operators that break in an explicit way the 
$SO(1,4)$ Lorentz symmetry  had been added. 
For simplicity, we will consider two explicit extensions of a 
$5D$ Wess-Zumino supersymmetric free model which, later, will be compactified 
using the orbifold $S^1/Z_2$. 
As we will discuss, our results show  different phenomenological consequences which depend on the extra dimensional 
parity assigned to the operators that violate the  $5D$ Lorentz symmetry.  Even though  the answer to our main question is 
on the positive at the $5D$ level, parity violating models turn out to be anomalous. Restoring SUSY 
in the lasts requires the addition of a $5D$ term that later generates a mass 
gap on $4D$ SUSY partners. This 
implies the explicit breaking of $4D$ SUSY, thus, connecting 
SUSY and Lorentz symmetry breaking. 
A close look at this model shows that, on the orbifold, the resulting model 
appears to be related to the one arising from the Scherk-Schwarz mechanism, where a twist on the boundary conditions is 
involved~\cite{ss,quiros}. This suggests that the last mechanism may have its 
origin on bulk Lorentz violation. 

This work has been organize with the following structure. In the second section we describe both the models and 
the operators that will be taken into account to achieve the explicit violation of Lorentz 
symmetry, on a general flat $5D$ space. 
The required conditions to preserve $5D$ SUSY in such models are also discussed 
there, without regard to the later effects produced by compactification.
Section three is dedicated to show the phenomenological consequences on the 
effective models that result out of the space compactification on the $S^1/Z_2$ orbifold. 
Finally, some further discussion and our conclusions are presented in section four. 
Two appendices have been added to provide 
our conventions for the construction of the models and the main features of the  
$S^1/Z_2$ orbifold, as well 
as to explain some basic,  but required,  details on the  $5D$ Wess-Zumino supersymmetric model.

%%%%%%%%%%%%%%%%%%%%%%%%%%%%%%%%
\section{Five-dimensional supersymmetric models}
%%%%%%%%%%%%%%%%%%%%%%%%%%%%%%%

In the work of Ref.~\citen{Colladay:1998fq}, the authors divide the 4D 
operators that explicitly break Lorentz symmetry in two kinds, 
those that are CPT invariant, and those that are not. 
However, as here we are interested only on  explicitly breaking the 5D Lorentz 
symmetry 
and retrieving the 4D symmetry for the effective models after compactification, 
there is no need to consider CPT violation (see Ref. \citen{Rizzo:2005um} for a discussion of CPT in this context).
Instead,  on 5D, a classification of operators can be made by separating 
those  that preserve  parity over the fifth
space dimension, from those which do not.

We will start by considering the $5D$ Wess-Zumino model  given by the Lagrangian
\begin{equation}
\mathcal{L}_{WZ}=\partial_M H_i^\dagger \partial^M H_i+
i\overline{\Psi}\Gamma^M\partial_M\Psi+F_i^\dagger F_i~, \label{WZ5Deq}
\end{equation} 
where $H_i$ are two scalar fields, $i= 1,2$; $\Psi$ is a Dirac spinor 
formed by 
two Weyl spinors, written as $\left(\lambda\ \bar{\chi}\right)^T$, and $F_i$ 
are two auxiliary 
fields (for further details on the model see \ref{wZ5D}).
In above equation  $M$ represents the five-dimensional indices, i.e. 
$M=\lbrace\mu, y\rbrace$, where we use 
$\mu=\lbrace 0,1,2,3\rbrace$ to refer to the standard four-dimensional indices,
whereas the fifth index is denoted as $y$. Gamma matrices $\Gamma^M=\lbrace 
\gamma^\mu, \gamma^y\rbrace$ are given in chiral representation as
\begin{equation}
\gamma^\mu=\left(\begin{array}{cc}
\mathbf{0}&\sigma^\mu\\
\bar{\sigma}^\mu&\mathbf{0}
\end{array}\right)~,\hspace{1cm}
\gamma^y=i\left(\begin{array}{cc}
-\mathbf{1}&\mathbf{0}\\\mathbf{0}&\mathbf{1}
\end{array}\right)~,
\end{equation}
where $\sigma^\mu=\left(\mathbf{1},\vec{\sigma}\right)$ and 
$\bar{\sigma}^\mu=\left(\mathbf{1},-\vec{\sigma}\right)$. Here, $\vec{\sigma}$ 
are 
the usual Pauli 
sigma matrices and the five-dimensional space-time is assumed to be flat with the 
Minkowski metric
\begin{equation}
\eta^{MN}=diag\left(1,-1,-1,-1,-1\right).
\end{equation}

%%%%%%%%%%%%%
\subsection{The Lorentz violating operators}
%%%%%%%%%%%%%
Given the fields contained on the Wess-Zumino model, the operators that violate 
$5D$ Lorentz symmetry, but preserve $4D$ Poincare invariance,  may 
only contain scalars ($H_i$) and fermion fields ($\Psi$). As mentioned previously, these 
operators can be separated in two sets, according to the $5D$ $Z_2$ parity. 
First, the operators that preserve $5D$ parity, that is, those that  are even under the simple mapping $Z_2:y\rightarrow -y$, 
are given by
\begin{eqnarray}
&&k_H\partial_y H^\dagger \partial^y H; \\
&&k_{\Psi}\overline{\Psi}\Gamma^y\partial_y\Psi .
\end{eqnarray}
On the other hand, there are also two operators where parity over the additional dimension is 
not preserved, which are given by
\begin{eqnarray}
&&i\alpha H^\dagger\partial_y H;\label{scacurr}\\
&&\beta\overline{\Psi}\Gamma^y\Psi. \label{fercurr}
\end{eqnarray}
In above, $k_\Psi$ and $k_H$ are dimensionless parameters, whereas the parameters $\alpha$ and 
$\beta$ have a mass dimension one. This mass dimensionality shall have important phenomenological consequences 
as we will seen in detail later.
We will discuss both the sets separately,  before moving into the corresponding effective $4D$ theory.
It is important to remark that the operators on Eqs.~(\ref{scacurr}) and (\ref{fercurr}) 
can be seen as the scalar and fermion fifth current component 
respectively, and they can interact with a gauge field, but this approach and 
its consequences after compactification are out of the scope of this work~\cite{ArkaniHamed:2001is}.

%%%%%%%%%%%%%%%%%%
\subsection{Parity preserving case}
%%%%%%%%%%%%%%%%%%%
Let us first consider those operators which respect $5D$ parity
to 
build an extended Lagrangian,  $\mathcal{L}=\mathcal{L}_{WZ}+\mathcal{L}_{\slashed{L}}$,
with an explicit rupture of $5D$ Lorentz symmetry, given 
by
\begin{equation}
\mathcal{L}_{\slashed{L}}=k_{H_i}\partial_y H_i^\dagger \partial^y H_i+
k_{\Psi}\overline{\Psi}\Gamma^y\partial_y\Psi. \label{preservedmodel}
\end{equation}
A direct calculation reveals that the model generated with the above Lagrangian is invariant, 
up to a total derivative, if the following extended supersymmetric transformations are 
considered
\begin{eqnarray}
\delta_{\xi} H_i&=&\sqrt{2}\epsilon_{ij}\overline{\xi}_j\Psi;\\
\delta_{\xi} \Psi&=&-i\sqrt{2}\epsilon_{ij}\left(\Gamma^M\partial_M+
k_\Psi \Gamma^y\partial_y\right)\xi_j H_i+\sqrt{2}\xi_i F_i;\\
\delta_{\xi}F_i&=&-i\sqrt{2}\,\overline{\xi}_i\left(\Gamma^M\partial_M+
k_\Psi \Gamma^y\partial_y\right)\Psi,
\end{eqnarray}
provided that, in order to close the algebra, the following restriction on the 
coupling parameters is impose,
\begin{equation}
k_{H_i}=k_\Psi^2+2k_\Psi.\label{susycond1}
\end{equation}

On the other hand, the commutator of two of these so defined SUSY 
transformations is given by
\begin{equation}
\left[\delta_\eta,\delta_\xi\right]\propto 
2i\overline{\eta}_i\left(\Gamma^M\partial_M+k_\Psi \Gamma^y\partial_y\right)\xi_i,
\end{equation}
where we can easily see  the translational operator dependence and that the 
supersymmetric transformations for an usual 5D Wess-Zumino model are recovered 
in the limit $k_\Psi\rightarrow0$ (see \ref{wZ5D}), as it would be expected.

To finalize this subsection, we note the possibility to make a redefinition for the spatial 
derivative in its fifth component
\begin{equation}
\partial_y\rightarrow\left(1+k_\Psi\right)\partial_y, \label{momentumdef}
\end{equation}
which allows to rewrite the Lagrangian for this model in the usual superfield formalism as
\begin{equation}
\mathcal{L}_1=\left.\overline{\Phi}_i \Phi_i\right|_D+
\left.\Phi_1\left(1+k_\Psi\right)\partial_y\Phi_2\right|_F+h.c.
\label{pinvl}
\end{equation}
In this equation, $\Phi_1$ and $\Phi_2$ are standard chiral superfields, as 
described in the \ref{wZ5D}. 

It is worth noticing that the whole effect 
imposed by the supersymmetry condition (\ref{susycond1}) is to reduce the overall factor in front of the fifth 
derivatives to a common one for all field components in the supermultiplet, 
as indicated by Eq.~(\ref{pinvl}). 
One may argue that such a 
constant factor can be removed by a simple rescaling of the fifth coordinate, such that  $ y \to  (1 + k_\Psi) y$. 
Such a mapping removes the overall factor on the fifth derivatives implied by 
Eq.~(\ref{momentumdef}), 
rendering an explicitly invariant theory, where Lorentz violating operators had been eliminated, and the extended 
Lagrangian is reduced to the standard Wess-Zumino model. One should then conclude that supersymmetry does 
require that the $5D$ free model be equivalent to a Lorentz invariant one. For a collection of free fields the same conclusion would arise 
even if the Lorentz violation parameters were non universal. That would be so, since in the absence of interactions, 
the action part of each free field can be treated separately. 
Thus, performing the proper mapping in each sector in an totally independent way 
would be straightforward. 
Nevertheless,  in a realistic model realization of this ideas more fields and interactions must be added. For instance, 
gauge and other required matter multiplets would be required. 
In the presence of interactions, as all other 
required kinetic terms will also involve derivatives on the fifth dimension, 
these will not allow  the complete removal of all Lorentz violating 
coefficients, unless they were universal. Therefore, Lorentz violation in supersymmetric $5D$ theories, implemented 
though parity preserving operators, would become a real effect provided its 
parameters are non universal. Notice however that maintaining supersymmetry in 
such interacting models still has to be established.
Considering this possibility and the fact that interactions are usually treated 
in the perturbative regime, 
we will proceed with our analysis of the effective compactified model in the next section without assuming the rescaling of the fifth coordinate. 
Nonetheless, we shall still comment on its effects on the Kaluza-Klein particle 
spectrum.

%%%%%%%%%%%%%%%%%%%%%%%%%%%%%%%%%%

\subsection{Parity violating case}
Next, let us consider once again the  Wess-Zumino model, but now with the 
addition 
of an $\mathcal{L}_{\slashed{L}}$ 
which takes into account the odd parity operators, those where parity over the fifth dimension is violated. Such a 
Lagrangian can be expressed as
\begin{equation}
\mathcal{L}_\slashed{L}=i\alpha H_i^\dagger\partial_y H_i+
\beta\overline{\Psi}\Gamma^y\Psi+h.c. \label{secondLag}
\end{equation}
The associated SUSY transformations for this model are then given by 
\begin{eqnarray}
\delta_\xi H_i&=&\sqrt{2}\epsilon_{ij}\overline{\xi}_j\Psi; \label{Scatrans2}\\
\delta_\xi \Psi&=&-i\sqrt{2}\epsilon_{ij}\left(\Gamma^M\partial_M+
i\alpha\Gamma^y\right)\xi_j H_i +\sqrt{2}\xi_i F_i\,;\\
\delta_\xi F_i&=&-i\sqrt{2}\, \overline{\xi}_i\left(\Gamma^M\partial_M+
i\alpha\Gamma^y\right)\Psi. \label{Ftrans2}
\end{eqnarray}
However, unlike what happens in the previous parity conserving case, 
here, in order to ensure the invariance under such deformed supersymmetric 
transformations, the parity violating terms in (\ref{secondLag}) do 
require to add a new $5D$ mass-like scalar term to the Lagrangian, given as
\begin{equation}
\mathcal{L}_m=-\alpha^2H_i^\dagger H_i~. \label{massterm}
\end{equation}
Furthermore, in addition to that, it becomes necessary to impose on the Lorentz violating parameters the restriction 
$\beta=-\alpha$, as it can be easily verified.

For this model, as for the parity preserving case,  it is also possible to make 
a redefinition for the fifth component of the momentum operator, as  
\begin{equation}
\partial'_y=\partial_y+i\alpha,\label{nonparityred}
\end{equation}
in order to simplify the transformations given in Eqs. 
(\ref{Scatrans2})-(\ref{Ftrans2}). 
Nonetheless, as it is clear, this redefinition does not conserve parity over 
the fifth dimension ($\partial'_y\neq \partial'_{-y}$). Of cou\-rse, this was 
actually expected, 
since $\alpha$ is precisely the  parameter associated to the parity 
violating  operators in  Eq.~(\ref{secondLag}).

The commutator for two of such supersymmetric transformations is now given by
\begin{equation}
\left[\delta_\eta,\delta_\xi\right]\propto 2\overline{\eta}_i\left(\Gamma^M\partial_M+
i\alpha\Gamma^y\right)\xi_i.
\end{equation}
From here, it is easy to see that for $\alpha\rightarrow 0$ the same features of the Wess-Zumino 
SUSY model would be recovered, as expected.  However, if $\alpha\neq 0$ there 
are important 
phenomenological implications that shall become explicit once the 
compactification is realized, as we will see later.

Similarly to the parity conservation model, the redefinition (\ref{nonparityred}) allows to write this model in 
the superfield formalism through the potentials
\begin{eqnarray}
K\left(\Phi_i\right)&=&\bar{\Phi}_i\Phi_i,\\
W\left(\Phi_1,\Phi_2\right)&=&\Phi_1\left(\partial_y+i\alpha\right)\Phi_2.
\end{eqnarray}
Again, the superpotential shows in an explicit way the Lorentz violation because the superfields $\Phi_1$ and $\Phi_2$ 
do have opposite $Z_2$ parities and  now it is not possible to 
hide this violation through a redefinition over the $y$-coordinate. Therefore, this specific model shows how Lorentz 
violation and supersymmetry could coexist in the frame of extra dimension models. 

Finally, we note that this model can be transformed into the Wess-Zumino  model (\ref{WZ5Deq}) through the next redefinition of the fields 
\begin{eqnarray}
H_i&\rightarrow& e^{-i\alpha y}H_i \nonumber\\
\Psi&\rightarrow& e^{-i\alpha y}\Psi\label{livtransf}\\
F_i&\rightarrow& e^{-i\alpha y}F_i ~.\nonumber
\end{eqnarray}
Consider for instance the fermion terms, which, under above redefinitions and 
the SUSY condition on the Lorentz violation parameters, become
\begin{eqnarray}
i\overline{\Psi}\Gamma^M\partial_M\Psi + \beta \overline{\Psi}\Gamma^y\Psi\rightarrow
i\overline{\Psi}\Gamma^\mu\partial_\mu\Psi +i\overline{\Psi}\Gamma^y\left(\partial_y-i\alpha\right)\Psi+\beta\overline{\Psi}\Gamma^y\Psi=
i\overline{\Psi}\Gamma^M\partial_M\Psi.\nonumber
\end{eqnarray}
A similar calculation confirms also the case for the scalar and the auxiliary field terms.
At a first glance, this observation suggests that the above field 
transformation can hide the Lorentz violation. 
This can be understood since the field transformations themselves are not in a $5D$ Lorentz covariant form, and 
therefore they do not commute  with Lorentz transformations. 
This leaves an open ambiguity. A SUSY Lorentz violating theory could be 
regarded as fundamental, 
which means that it would be derived at such from first principles, from String 
theory for instance, 
with the Lorentz violation parameters associated to the vacuum expectation values of 
some bulk background fields. 
But then, the fields in the transformed frame where Lorentz violation has been hidden should have odd $5D$ 
Lorentz transformations to compensate for the properties of the local phase factor. 
The opposite situation is also possible. One may introduce an apparent Lorentz violation 
in a SUSY Lorentz conserving theory through the inverse transformations of those given in Eq.~(\ref{livtransf}), 
in which case, the effect should be regarded as non physical. 
To disentangle the ambiguity one would have to relay on the implications that the model has at the effective level, 
once the $5D$ theory is sited on a compactified space, as we will discuss below.

%%%%%%%%%%%%
\section{Effective 4D models}
%%%%%%%%%%%%%
As it is usual for extra dimensional models, to get a sensible four dimensional 
effective theory it 
is necessary to perform a compactification of the extra space. There are many ways to achieve this. 
For our following discussion, to be specific, we shall consider a compact fifth dimension on the orbifold 
$S^1/Z_2$ of radius $R$ (see  \ref{orbifoldapp} for details). 
We will also assume that the Lorentz violating 
frame should be regarded as fundamental, and then discuss the implications of the model on the orbifolded theory. 
That means that, along the analysis, all bulk fields will be regarded to comply 
with the general periodicity condition
$\phi(y+2\pi R)= \phi(y)$.

As it is mentioned in the \ref{orbifoldapp}, it is possible to assign a specific Kaluza-Klein
(KK) decomposition of the fields in order to make the $Z_2$ parity explicit. This can be written for the superfields as
\begin{eqnarray}
\Phi_1&=&\frac{1}{\sqrt{\pi R}}\Phi_1^{\left(0\right)}\left(x^\mu\right)+
\sqrt{\frac{2}{\pi R}}\sum_{n=1}\Phi_1^{\left(n\right)}\left(x^\mu\right)\cos\left(\frac{ny}{R}\right)~,\\
\Phi_2&=&\sqrt{\frac{2}{\pi R}}\sum_{n=1}\Phi_2^{\left(n\right)}\left(x^\mu\right)\sin\left(\frac{ny}{R}\right)~.
\label{kkmodes}
\end{eqnarray}
It is worth noticing that the chosen parities shall project out half of the 
KK modes, leaving, however, at each KK level the right configuration of fields 
as to conform a standard $4D$ chiral superfield model. This means to say that, 
in a standard $5D$ Wess-Zumino theory with the same boundary conditions, 
the $4D$ $N=2$ SUSY described by the field content will be broken, but  full 
$N=1$ SUSY field representations will be preserved by the chosen 
compatification  at each level of the KK tower.
Next, we will address the effects introduced in the effective theory by the $5D$ 
Lorentz 
violating operators, that is, for each of the models presented above.

First, for the parity preserving model, the zero mode on the Kaluza-Klein tower 
contains 
only the massless fields $H_{10}$,  
$\lambda_0$ and $F_{10}$. However, the excited modes $\left(n\geq 1\right)$ 
are susceptible to the Lorentz violating terms. As it is easy to see, the terms 
in Eq.~(\ref{preservedmodel}) become mass terms in the effective $4D$ theory, 
and thus, the mass for the 
$H_{1n}$, $H_{2n}$ and $\Psi_n$ fields shall differ from the Lorentz invariant Wess-Zumino 
model. In this case KK masses are given by
\begin{equation}
m_n^2=\left(\frac{1+k_\Psi}{R}\right)^2n^2,
\end{equation}
where $k_\Psi$ can take any real value without creating tachionic states since 
the minimum value of the function $k_\Psi^2+2k_\Psi$ is minus one. 
Last spectrum was actually expected due to the redefinition given in 
Eq.~(\ref{momentumdef}) that allows to hide the Lorentz 
violation terms to have a Lorentz invariant model. 
Notice that, as one would also expect, the effect of the bulk Lorentz 
violating terms in the effective theory is indeed equivalent to a rescaling of the 
size of the compact space. In a model with more than one family, the Lorentz 
violating parameter $k_\Psi$ can be family dependent. Therefore, the direct 
conclusion is that each particle KK tower would have associated to it a different effective 
size of the compact space. This particle dependent mass gap is a distinguishing 
feature of the bulk Lorentz violating operators, 
since in a truly Lorentz invariant theory the mass gap should be universal.

It is also straightforward to check that the $N=1$ SUSY remains unbroken. 
As a matter of fact, upon compactification, our first model can  be rewritten in superfield 
formalism as
\begin{eqnarray}
\mathcal{L}_{eff}&=&\sum_{n=1}^\infty\left[\left.\Phi_{in}^\dagger\Phi_{in}\right|_D+
\left.\frac{n}{R}\left(1+k_\Psi\right)\Phi_{1n}\Phi_{2n}\right|_F\right]\nonumber\\&&+
\left.\Phi_0^\dagger\Phi_0\right|_D+h.c.
\end{eqnarray}
where the KK chiral superfields are given by 
\begin{eqnarray}
\Phi_{0}&=&H_{10}+\sqrt{2}\lambda_0+\theta^2F_{10};\\
\Phi_{1n}&=&H_{1n}+\sqrt{2}\lambda_n+\theta^2 F_{1n};\\
\Phi_{2n}&=&H_{2n}+\sqrt{2}\chi_n+\theta^2 F_{2n}.
\end{eqnarray}

On the other hand, in our second model, the parity violating case,  compactification leads to a great difference 
relative to the  first model, because the terms which violate parity on the extra dimension, in Eq.~ (\ref{secondLag}), 
do not have any  contribution on the effective model, since being odd their integral over the extra dimension simply vanishes. 
However,  that is not the case for the additional term in Eq.~(\ref{massterm}). 
At the level of the effective model, as it is easy to see, this  term becomes an
universal mass term for all scalar KK modes. This is particularly interesting 
for the zero  mode level theory, because it means a mass term for the scalar field which has 
no equivalent  for the fermion field. Thus, we get the zero mode spectrum
\begin{eqnarray}
m_{H_{10}}^2&=&\alpha^2;\\
m_{\lambda_0}^2&=&0.
\end{eqnarray}
This, of course, represents a mass gap that evidences SUSY breaking on the 
$4D$ effective theory. It is worth stressing that the mass-like term in 
Eq.~(\ref{massterm}) was required to restore SUSY at the five dimensional 
level, due to the explicit breaking of $5D$ Lorentz invariance. 
Nonetheless, it is precisely 
this same term which breaks SUSY in the effective KK theory, where $4D$ Lorentz 
invariance holds. 
Interestingly enough, the term in consideration is 
precisely what is called a soft-breaking term. Furthermore, 
we notice that the emergence of the mass gap is actually independent of 
the chosen compactification, since there is no term in the Lagrangian that may 
generate an equivalent mass for the zero mode fermions.

A mass gap appears of course in the same way for all the excited modes, for which 
we get
\begin{eqnarray}
m_{H_{1n}}^2&=&m_{H_{2n}}^2=\alpha^2+\frac{n^2}{R^2};\\
m_{\lambda_n}^2&=&m_{\chi_{n}}^2=\frac{n^2}{R^2}.\label{pvspect}
\end{eqnarray}
This result shows that SUSY is broken in all the levels of the $4D$ effective 
theory.  Although this is a tree level calculation, at least for this toy model 
the mass gap is valid at any order, since 
we have not consider interactions so far, which could incorporate non universal 
mass radiative corrections.
Also, we notice that there is no restriction to the value for the $\alpha$ parameter in this case, 
besides that it should be a real number. 

 %%%%%%%%%%%%%%%%%%%%%%%%%%%%
By looking at the mass spectrum, we notice that this resembles the mass 
structure of KK towers in 
$5D$ models where supersymmetry is broken by the Scherk-Schwarz (SS) 
mechanism~\cite{ss,quiros}.  
The SS mechanism breaks supersymmetry 
by imposing on the bulk fields the non trivial periodic condition  
$\phi(y+2\pi R) = e^{i2\pi qT}\phi(y)$, where $T$ must be a non trivial 
generator
of a global symmetry of the $5D$ theory and $q$ its associated charge. 
Notice, however, 
that the mass spectrum in Eq.~(\ref{pvspect}) is shifted by an universal  radius independent term, whereas in 
the former the twist usually appears as a shift on the KK index, of the form $n\rightarrow n+q$.
The twisted condition implies that the fields can not longer be expressed 
in the standard KK mode expansion described by Eq.~(\ref{kkmodes}). Instead, the 
twisted fields are expressed as 
\begin{equation}
\label{twisted}
\phi(y) = e^{iqTy/R}\tilde\phi(y)~,
\end{equation}
where $\tilde\phi$ stands for a new field with standard periodicity conditions, 
$\tilde\phi(y+2\pi R) = \tilde\phi(y)$.
Obviously, $\tilde\phi$ does expand in the usual KK modes, according with its own parity.
When building models such a twist is usually assumed ad hoc. No further physical reason is claimed beneath the mechanism. 
Interestingly enough, the twisted transformation (\ref{twisted}) is actually equivalent to the ones given in 
Eq.~(\ref{livtransf})  that are used to hide Lorentz violation in the parity violating model. 
Furthermore, by considering our previous discussion, where we assumed that the 
fields in the Lorentz violating mode have standard periodicity conditions, it is straightforward to see that the transformation (\ref{livtransf}) 
actually maps our model into a frame  where the SS boundary conditions hold for all fields, but where the $T$ generator is just the identity with 
an universal charge identified as $q = \alpha R$. This is of course distinctive from the standar implementation of the SS mechanism.
Therefore, we can argue that our model actually incorporates a SS-like twist in a natural way, 
and that the last has a very well identified origin on 
the $5D$ supersymmetric Lorentz violation produced by non parity conserving operators.
This results may unveil a deeper connection among fundamental Lorentz violation in supersymmetric higher dimensional theories and the SS mechanism. 

%%%%%%%%%%%%%%%%%%%%%%%%%%%%%

%%%%%%%%%%%%%%
\section{Concluding remarks}
%%%%%%%%%%%%%%

In this work we have presented a study of an extended
Wess-Zumino supersymmetric model on five space-time dimensions, where
$5D$ Lorentz symmetry is violated in an explicit way. 
All possible field operators, up to mass dimension five,  with such an explicit 
violation had been considered. These operators can be classified as even or odd 
under the $Z_2$ fifth dimensional parity. Thus, we have  consi\-dered the 
two most general extended models:  the one where $Z_2$ parity is 
conserved and the one where parity over the additional dimension is not 
present.  
It has been argue that, in both the cases,  supersymmetry does still remain as a symmetry of the $5D$ 
theory under a set of deformed field transformations, provided 
some specific conditions on the Lorentz violation parameters are met. 
However, for the non  parity preserving models, restoring supersymmetry does 
also 
require the addition of an universal scalar mass term. 
It is worth mentioning that the SUSY transformations for both 
the models can be rewritten as those of a supersymmetric Wess-Zumino model 
through an ad hoc redefinition on the fifth momentum 
component.

In both the cases, the Lorentz violating terms can be hidden through some 
appropriate transformations, making the theory to appear as a Lorentz invariant 
one. 
Either by scaling the extra dimension, for the parity even theory, or by performing a non Lorentz covariant transformation in the parity odd model.
However, since any extra dimensional theory should consider the extra space to be compact,  
the effects of such transformations are  reflected as a 
deformation of the boundary conditions. Due to this, it seems likely that  
a simpler interpretation and treatment of the effective model would arise in the  Lorentz violating frame.

As a matter of fact, after compactification is incorporated  in the parity  preserving model, the only 
indication about the existence of the $5D$ Lorentz violating terms is a 
shifting of the squared masses for all KK excited 
modes. 
The shifting comes out to be  proportional to the squared KK number, and it is 
consistent with the coordinate rescaling that hides the Lorentz violating 
terms in the bulk. The rescaling is in general expected to be particle dependent, and thus, different particles will show different shifting. 
This is a distinctive feature of the model with respect to the truly Lorentz 
invariant free theories, where the KK spectrum is expected to be universal.
The zero mode fields, however, do remain massless. 
In this case, $N=1$ SUSY is preserved after compactification on the $S^1/Z_2$ 
orbifold, and thus, it is 
possible to rewrite the effective $4D$ theory in terms of an infinity tower of chiral superfields.
Here a comment is in order. Along our discussion we have only analyzed a free particle model. The introduction of interactions in 
the presence of $5D$ SUSY and Lorentz violation is an issue that is pending to be analyzed. 
The restoration of SUSY may be troublesome, since in the rescaled frame, where Lorentz violation is hidden, different fields would appear to have different 
periodicity, and that could make difficult for SUSY to prevail at the effective theory, which, yet, can be a desirable feature.

On the other hand, the non  $Z_2$ parity preserving model turns out to be 
the most interesting, at least theoretically. The additional scalar mass 
term required by the deformed SUSY, after compactification, becomes a supersymmetric soft breaking term for 
the KK tower that affects even the zero mode. Interestingly enough, the 
effective $4D$ theory of this model turns out to be non supersymmetric, even 
though its $5D$ parent it is so. Universality, and non derivative nature of the 
mentioned mass term, indicates that this results is independent of the choice 
for the compact space.
Another point to remark is that the compactification process over the $S^1/Z_2$ 
orbifold, that we have considered for our analysis,  cancels out both the 
operators which violate $5D$ Lorentz symmetry. And  so, these operators do not 
have any implication whatsoever in the effective four dimensional  
model. They are canceled out after integrating over the extra dimension
due to their explicit violation of $Z_2$ parity. 

We have shown that the transformation that may be used to hide the Lorentz 
violating operators in the bulk
do actually transforms the fields into some with similar properties as those 
required  by 
the Scherk-Schwarz mechanism. In this last frame, the field boundary conditions  acquire a natural twist, with a 
charge proportional to the Lorentz violating parameter of the theory. That may explain the breaking of SUSY on the effective $4D$ theory, 
but also suggest that bulk Lorentz violation may be the physics beneath the SS mechanism. Nonetheless, in this realization of the SS mechanism, 
the KK spectrum is quite different to the one usually obtained when non trivial symmetries are used (see for intance ref. \citen{quiros}). 
We think this idea does deserve further study.

%%%%%%%%%%%%%%%%%%%%%%%%%
Finally, the results 
we have presented  may also suggest a link for susy soft breaking terms to parity 
and Lorentz invariance violation in 5D models.  The realization 
of the this idea  on more realistic models may deserve further 
attention too. 
Extending our results by  taking into account
interaction between fields, either by adding them in the  superpotential or by 
the introduction of gauge fields, 
which would make the model more realistic, seems as an interesting possibility. 
In particular, it would be  quite interesting to address the question of 
a possible connection among the parity violating operators and the soft breaking terms needed in any
realistic supersymmetric model, as in the MSSM.

\section*{Acknowledgments}
We would like to thank one of the referees for some very useful comments. 
This work was supported in part by CONACYT, Mexico, under grant No. 237004.

\appendix

%%%%%%%%%%%%%%%%%%%
\section{5D supersymmetric Wess-Zumino model} \label{wZ5D}
%%%%%%%%%%%%%%%%%%%
The first supersymmetric field theory in the context of $4D$ was presented by 
Wess and Zumino~\cite{Wess:1973kz}.  It is straightforward to promote such 
model to  $5D$. For that, we just have to take into account two scalar fields, 
$H_i$, a  Dirac spinor, $\Psi$ and two auxiliary fields, $F_i$, for $i=1,2$.
The Lagrangian for such a model  is then given by
\begin{equation}
\mathcal{L}_{WZ}=\partial_M H_i^\dagger \partial^M 
H_i+i\overline{\Psi}\Gamma^M\partial_M\Psi+F_i
^\dagger F_i,
\end{equation}
where $M=0,1,2,3,5$.
It is easy to see that this last Lagrangian is invariant, up to a total 
derivative, under the following SUSY transformations
\begin{eqnarray}
\delta_{\xi} H_i&=&\sqrt{2}\epsilon_{ij}\overline{\xi}_j\Psi;\\
\delta_{\xi} \Psi&=&-i\sqrt{2}\epsilon_{ij}\Gamma^M\partial_M\xi_j 
H_i+\sqrt{2}\xi_i F_i\,;\\
\delta_{\xi}F_i&=&-i\sqrt{2}\,\overline{\xi}_i\Gamma^M\partial_M\Psi.
\end{eqnarray}
%\end{appendix}
Here, $\epsilon_{ij}$ is the antisymmetric tensor 
($\epsilon_{21}=1=-\epsilon_{12}$) and $\xi_i$ 
is a symplectic Majorana spinor which represents a constant parameter for the 
supersymmetric transformations.

As a remainder to the reader, a symplectic Majorana spinor satisfy the 
relation 
$\xi_i=\epsilon_{ij}C\overline{\xi}_j^T$, where $C$ is the charge conjugation 
operator. They can be written through two Weyl spinors as
\begin{equation}
\xi_1=\left(\begin{array}{c}
\varepsilon\\\bar{\eta}
\end{array}\right);\hspace{1cm}\xi_2=\left(\begin{array}{c}
\eta\\-\bar{\varepsilon}
\end{array}\right).
\end{equation}
It is thanks to these spinors that it is possible to write the superalgebra as
\begin{equation}
\lbrace Q_i,Q_j\rbrace=2\Gamma^M P_M \delta_{ij}.
\end{equation} 

Similar to the 4D case, the 5D Wess-Zumino model can be written in standard 
superfield notation 
\cite{ArkaniHamed:2001tb}, by considering the following two chiral superfields
\begin{eqnarray}
\Phi_1=H_1+\sqrt{2}\theta\lambda+\theta^2 F_1;\\
\Phi_2=H_2+\sqrt{2}\theta\chi+\theta^2 F_2,
\end{eqnarray}
in terms of which the action for the Wess-Zumino Lagrangian can be written as 
\begin{equation}
S_{WZ}^5=\int d^4 x dy\left[\int 
d^4\theta\left(\Phi_1^\dagger\Phi_1+\Phi_2^\dagger\Phi_2\right)+\int 
d^2\theta\Phi_1\partial_y\Phi_2+h.c.\right]
\end{equation}
assuming $\Psi=\left(\lambda,\ \bar{\chi}\right)^T$.

%%%%%%%%%%%%%%%%
\section{The compactification and the orbifold $S^1/Z_2$ }\label{orbifoldapp}
%%%%%%%%%%%%%%%%
Effective $4D$ theory  arises from an 
extra dimensional space-time model when a compactification of the extra 
dimensions is imposed. This compactification process dictates many of the 
geometric properties of the effective model and, in many 
cases, it has an explicit influence over the couplings between the 
effective fields.

In $5D$ models the most easy way to achieve this compactification process involves to consider 
the extra dimension to be a circle of radius $R$. This selection 
allow us to give a explicit description to the fields in terms of a Fourier 
expansion,  
\begin{equation}
\phi\left(x^\mu,y\right)\sim\sum_{n}\phi_n\left(x^\mu\right)e^{i \frac{n}{R} y}. 
\label{compactcircle}
\end{equation}
where the $n-th$ mode, $\phi_n$, corresponds to one of the cyclic modes that 
moves around the extra dimension.
As an example, consider a free scalar field, $\phi$,  whose 
$5D$ Lagrangian is given  by
\begin{equation}
\mathcal{L}=\partial_M \phi^\dagger\partial^M\phi\,.
\end{equation}
The compactification on the circle then leads to the effective $4D$ model 
described by the Lagrangian
\begin{eqnarray}
\mathcal{L}_{eff}&=&\int_{-\pi R}^{\pi R}\partial_M \phi^\dagger\partial^M\phi dy\\
&=&\sum_{n=0}^{\infty}\left(\partial_\mu \phi^\dagger_n\partial^\mu \phi_n-
\frac{n^2}{R^2}\left|\phi_n\right|^2\right),
\end{eqnarray}
where we can see that  the compactification provides, from a purely 
$5D$ massless model, an infinite set of scalar Kaluza-Klein field modes, where 
only one of them, the zero mode, remains massless.

Compactification on a circle, however, does not assign explicit extra 
dimensional parity eigenvalues  to  the fields. This can be done, if the 
discrete group $Z_2$ over the 
circle is included.
This amounts to include an additional identification of opposite points along 
the circle, those associated by the transformation $Z_2:y\rightarrow -y$. With 
this, it becomes  now 
possible to label the fields with the parity eigenvalues $\pm 1$. That means to 
consider effective field expansions in the two following possible forms
\begin{eqnarray}
\phi_+\left(x^\mu,y\right)&\sim&\sum_{n}\phi_n\left(x^\mu\right)\cos\left(\frac{
n }{R}y\right),\\
\phi_-\left(x^\mu,y\right)&\sim&\sum_{n}\phi_n\left(x^\mu\right)\sin\left( 
\frac{n}{R} y\right).
\end{eqnarray}
This last  decomposition for the fields is the right expansion to consider when 
compactification is done over the $S^1/Z_2$ orbifold. Notice that the effect is 
the absence of half of the KK  modes with respect to (\ref{compactcircle}).

%\bibliographystyle{unsrt} 
%\bibliography{referencias}

\begin{thebibliography}{X}

%\cite{Coleman:1967ad}
\bibitem{coleman1988aspects}
Coleman Sidney, \emph{Aspects of symmetry: selected Erice lectures}, Cambridge University Press(1988) 


%\cite{Martin:1997ns}
\bibitem{Martin:1997ns} 
  S.~P.~Martin,
  %``A Supersymmetry primer,''
  Adv.\ Ser.\ Direct.\ High Energy Phys.\  {\bf 21}, 1 (2010)
  [Adv.\ Ser.\ Direct.\ High Energy Phys.\  {\bf 18}, 1 (1998)]
  %doi:$10.1142/9789812839657\_0001,\ 10.1142/9789814307505\_0001$
  [hep-ph/9709356].
  %%CITATION = doi:10.1142/9789812839657_0001, 10.1142/9789814307505_0001;%%
  %2740 citations counted in INSPIRE as of 03 Aug 2016  

\bibitem{XD}
N. Arkani-Hamed, S. Dimopoulos and G. Dvali, Phys. Lett. B {\bf 429}, 263 (1998);
I. Antoniadis, et al., Phys. Lett. B {\bf 436}, 257 (1998);
I. Antoniadis, S. Dimopoulos, G. Dvali, Nucl. Phys. B {\bf 516}, 70 (1998).
  
 
  
  %\cite{PerezLorenzana:2005iv}
\bibitem{PerezLorenzana:2005iv} 
For a review see
  A.~Perez-Lorenzana,
  %``An Introduction to extra dimensions,''
  J.\ Phys.\ Conf.\ Ser.\  {\bf 18}, 224 (2005)
 % doi:10.1088/1742-6596/18/1/006
  [hep-ph/0503177].
  %%CITATION = doi:10.1088/1742-6596/18/1/006;%%
  %66 citations counted in INSPIRE as of 07 Jul 2016  
    
  
%\cite{VanProeyen:1999ni}
\bibitem{VanProeyen:1999ni} 
  A.~Van Proeyen,
  %``Tools for supersymmetry,''
  Ann.\ U.\ Craiova Phys.\  {\bf 9}, no. I, 1 (1999)
  [hep-th/9910030].
  %%CITATION = HEP-TH/9910030;%%
  %179 citations counted in INSPIRE as of 03 Aug 2016   
   
%\cite{Marti:2001iw}
\bibitem{Marti:2001iw} 
  D.~Marti and A.~Pomarol, 
  %``Supersymmetric theories with compact extra dimensions in N=1 superfields,''
  Phys.  Rev.  D {\bf 64}, 105025 (2001)
%  doi:10.1103/PhysRevD.64.105025
  [hep-th/0106256].
  %%CITATION = doi:10.1103/PhysRevD.64.105025;%%
  %211 citations counted in INSPIRE as of 22 Nov 2016   
   
%\cite{Berger:2001rm}
\bibitem{Berger:2001rm} 
  M.~S.~Berger and V.~A.~Kostelecky,
  %``Supersymmetry and Lorentz violation,''
  Phys.\ Rev.\ D {\bf 65}, 091701 (2002)
  %doi:10.1103/PhysRevD.65.091701
  [hep-th/0112243].
  %%CITATION = doi:10.1103/PhysRevD.65.091701;%%
  %116 citations counted in INSPIRE as of 07 Jul 2016
  
\bibitem{ss}
J. Scherk, J.H. Schwarz, Nucl. Phys. B {\bf 153}, 61 (1979); Phys. Lett. B {\bf 82}, 60 (1979); 
E. Cremmer, J. Scherk, J.H. Schwarz, Phys. Lett. B {\bf 84}, 83 (1979);
P. Fayet, Phys. Lett. B {\bf 159}, 121 (1985); Nucl. Phys. B {\bf 263}, 649 (1986).

\bibitem{quiros}
See for instance: A. Pomarol and M. Quiros. 
Phys. Lett. B {\bf 438}, 255  (1998).
  
  
  
%\cite{Colladay:1996iz}
\bibitem{Colladay:1996iz} 
  D.~Colladay and V.~A.~Kostelecky,
  %``CPT violation and the standard model,''
  Phys.\ Rev.\ D {\bf 55}, 6760 (1997)
  %doi:10.1103/PhysRevD.55.6760
  [hep-ph/9703464].
  %%CITATION = doi:10.1103/PhysRevD.55.6760;%%
  %1128 citations counted in INSPIRE as of 07 Jul 2016
  
%\cite{Kostelecky:1988zi}
\bibitem{Kostelecky:1988zi} 
  V.~A.~Kostelecky and S.~Samuel,
  %``Spontaneous Breaking of Lorentz Symmetry in String Theory,''
  Phys.\ Rev.\ D {\bf 39}, 683 (1989).
  %doi:10.1103/PhysRevD.39.683
  %%CITATION = doi:10.1103/PhysRevD.39.683;%%
  %862 citations counted in INSPIRE as of 03 Dec 2016  
  
  %\cite{Wess:1973kz}
\bibitem{Wess:1973kz} 
  J.~Wess and B.~Zumino,
  %``A Lagrangian Model Invariant Under Supergauge Transformations,''
  Phys.\ Lett.\ B {\bf 49}, 52 (1974).
  %doi:10.1016/0370-2693(74)90578-4
  %%CITATION = doi:10.1016/0370-2693(74)90578-4;%%
  %1357 citations counted in INSPIRE as of 07 Jul 2016
  
  %\cite{Kostelecky:2008ts}
\bibitem{Kostelecky:2008ts} 
  V.~A.~Kostelecky and N.~Russell,
  %``Data Tables for Lorentz and CPT Violation,''
  Rev.\ Mod.\ Phys.\  {\bf 83}, 11 (2011)
  %doi:10.1103/RevModPhys.83.11
  [arXiv:0801.0287 [hep-ph]].
  %%CITATION = doi:10.1103/RevModPhys.83.11;%%
  %585 citations counted in INSPIRE as of 07 Jul 2016
  
  
\bibitem{petrov}
For similar models see for instance:
C. F. Farias, A. C. Lehum, J. R. Nascimento, A. Yu. Petrov,
Phys. Rev. D {\bf 86}, 065035 (2012);
A. C. Lehum, J. R. Nascimento, A. Yu. Petrov, A. J. da Silva,
Phys. Rev. {\bf D88}, 045022 (2013).
  
  
\bibitem{Rizzo:2005um} 
  T.~G.~Rizzo,
  %``Lorentz violation in extra dimensions,''
  JHEP {\bf 0509}, 036 (2005)
  %doi:10.1088/1126-6708/2005/09/036
  [hep-ph/0506056].
  %%CITATION = doi:10.1088/1126-6708/2005/09/036;%%
  %18 citations counted in INSPIRE as of 07 Jul 2016
  
  
  %\cite{Colladay:1998fq}
\bibitem{Colladay:1998fq} 
  D.~Colladay and V.~A.~Kostelecky,
  %``Lorentz violating extension of the standard model,''
  Phys.\ Rev.\ D {\bf 58}, 116002 (1998)
  %doi:10.1103/PhysRevD.58.116002
  [hep-ph/9809521].
  %%CITATION = doi:10.1103/PhysRevD.58.116002;%%
  %1445 citations counted in INSPIRE as of 03 Aug 2016
  
 %\cite{ArkaniHamed:2001is}
\bibitem{ArkaniHamed:2001is} 
  N.~Arkani-Hamed, A.~G.~Cohen and H.~Georgi,
  %``Anomalies on orbifolds,''
  Phys.\ Lett.\ B {\bf 516}, 395 (2001)
  %doi:10.1016/S0370-2693(01)00946-7
  [hep-th/0103135].
  %%CITATION = doi:10.1016/S0370-2693(01)00946-7;%%
  %137 citations counted in INSPIRE as of 03 Aug 2016 
  
  
  %\cite{ArkaniHamed:2001tb}
\bibitem{ArkaniHamed:2001tb} 
  N.~Arkani-Hamed, T.~Gregoire and J.~G.~Wacker,
  %``Higher dimensional supersymmetry in 4-D superspace,''
  JHEP {\bf 0203}, 055 (2002)
  %doi:10.1088/1126-6708/2002/03/055
  [hep-th/0101233].
  %%CITATION = doi:10.1088/1126-6708/2002/03/055;%%
  %285 citations counted in INSPIRE as of 07 Jul 2016


\bibitem{quiros2}
See for instance:
 A. Delgado, G. von Gersdorff and M. Quiros, JHEP12 (2002).
  
  
  \end{thebibliography}
\end{document}